


\font\bigbold=cmbx12
\font\eightrm=cmr8
\font\sixrm=cmr6
\font\fiverm=cmr5
\font\eightbf=cmbx8
\font\sixbf=cmbx6
\font\fivebf=cmbx5
\font\eighti=cmmi8  \skewchar\eighti='177
\font\sixi=cmmi6    \skewchar\sixi='177
\font\fivei=cmmi5
\font\eightsy=cmsy8 \skewchar\eightsy='60
\font\sixsy=cmsy6   \skewchar\sixsy='60
\font\fivesy=cmsy5
\font\eightit=cmti8
\font\eightsl=cmsl8
\font\eighttt=cmtt8
\font\tenfrak=eufm10
\font\sevenfrak=eufm7
\font\fivefrak=eufm5
\font\tenbb=msbm10
\font\sevenbb=msbm7
\font\fivebb=msbm5
\font\tensmc=cmcsc10
\font\tencmmib=cmmib10  \skewchar\tencmmib='177
\font\sevencmmib=cmmib10 at 7pt \skewchar\sevencmmib='177
\font\fivecmmib=cmmib10 at 5pt \skewchar\fivecmmib='177

\newfam\bbfam
\textfont\bbfam=\tenbb
\scriptfont\bbfam=\sevenbb
\scriptscriptfont\bbfam=\fivebb

\newfam\frakfam
\textfont\frakfam=\tenfrak
\scriptfont\frakfam=\sevenfrak
\scriptscriptfont\frakfam=\fivefrak

\newfam\cmmibfam
\textfont\cmmibfam=\tencmmib
\scriptfont\cmmibfam=\sevencmmib
\scriptscriptfont\cmmibfam=\fivecmmib
\def\bold#1{\fam\cmmibfam\relax#1}


\def\eightpoint{%
\textfont0=\eightrm   \scriptfont0=\sixrm
\scriptscriptfont0=\fiverm  \def\rm{\fam0\eightrm}%
\textfont1=\eighti   \scriptfont1=\sixi
\scriptscriptfont1=\fivei  \def\oldstyle{\fam1\eighti}%
\textfont2=\eightsy   \scriptfont2=\sixsy
\scriptscriptfont2=\fivesy
\textfont\itfam=\eightit  \def\it{\fam\itfam\eightit}%
\textfont\slfam=\eightsl  \def\sl{\fam\slfam\eightsl}%
\textfont\ttfam=\eighttt  \def\tt{\fam\ttfam\eighttt}%
\textfont\bffam=\eightbf   \scriptfont\bffam=\sixbf
\scriptscriptfont\bffam=\fivebf  \def\bf{\fam\bffam\eightbf}%
\abovedisplayskip=9pt plus 2pt minus 6pt
\belowdisplayskip=\abovedisplayskip
\abovedisplayshortskip=0pt plus 2pt
\belowdisplayshortskip=5pt plus2pt minus 3pt
\smallskipamount=2pt plus 1pt minus 1pt
\medskipamount=4pt plus 2pt minus 2pt
\bigskipamount=9pt plus4pt minus 4pt
\setbox\strutbox=\hbox{\vrule height 7pt depth 2pt width 0pt}%
\normalbaselineskip=9pt \normalbaselines
\rm}


\def\pagewidth#1{\hsize= #1}
\def\pageheight#1{\vsize= #1}
\def\hcorrection#1{\advance\hoffset by #1}
\def\vcorrection#1{\advance\voffset by #1}

\newcount\notenumber  \notenumber=1              
\newif\iftitlepage   \titlepagetrue              
\newtoks\titlepagefoot     \titlepagefoot={\hfil}
\newtoks\otherpagesfoot    \otherpagesfoot={\hfil\tenrm\folio\hfil}
\footline={\iftitlepage\the\titlepagefoot\global\titlepagefalse
           \else\the\otherpagesfoot\fi}

\def\abstract#1{{\parindent=30pt\narrower\noindent\eightpoint\openup
2pt #1\par}}
\def\smc{\tensmc}


\def\note#1{\unskip\footnote{$^{\the\notenumber}$}
{\eightpoint\openup 1pt
#1}\global\advance\notenumber by 1}

\def\frac#1#2{{#1\over#2}}

\def\tfrac#1#2{{\textstyle{#1\over#2}}}
\def\({\left(}
\def\){\right)}
\def\<{\langle}
\def\>{\rangle}
\def\2pd#1#2#3{\frac{\partial^2#1}{\partial#2\partial#3}}

\def\sqr#1#2{{\vcenter{\vbox{\hrule height.#2pt
        \hbox{\vrule width.#2pt height#1pt \kern#1pt
           \vrule width.#2pt}
        \hrule height.#2pt}}}}

\def\ni{\noindent}
\def\lqq{\lq\lq}
\def\rqq{\rq\rq}


\global\newcount\secno \global\secno=0
\global\newcount\meqno \global\meqno=1
\global\newcount\appno \global\appno=0
\newwrite\eqmac
\def\romappno{\ifcase\appno\or A\or B\or C\or D\or E\or F\or G\or H
\or I\or J\or K\or L\or M\or N\or O\or P\or Q\or R\or S\or T\or U\or
V\or W\or X\or Y\or Z\fi}
\def\eqn#1{
        \ifnum\secno>0
            \eqno(\the\secno.\the\meqno)\xdef#1{\the\secno.\the\meqno}
          \else\ifnum\appno>0
            \eqno({\rm\romappno}.\the\meqno)\xdef#1{{\rm\romappno}.
               \the\meqno}
          \else
            \eqno(\the\meqno)\xdef#1{\the\meqno}
          \fi
        \fi
\global\advance\meqno by1 }

\def\eqnn#1{
        \ifnum\secno>0
            (\the\secno.\the\meqno)\xdef#1{\the\secno.\the\meqno}
          \else\ifnum\appno>0
            \eqno({\rm\romappno}.\the\meqno)\xdef#1{{\rm\romappno}.
                \the\meqno}
          \else
            (\the\meqno)\xdef#1{\the\meqno}
          \fi
        \fi
\global\advance\meqno by1 }

\global\newcount\refno
\global\refno=1 \newwrite\reffile
\newwrite\refmac
\newlinechar=`\^^J
\def\ref#1#2{\the\refno\nref#1{#2}}
\def\nref#1#2{\xdef#1{\the\refno}
\ifnum\refno=1\immediate\openout\reffile=refs.tmp\fi
\immediate\write\reffile{
     \noexpand\item{[\noexpand#1]\ }#2\noexpand\nobreak.}
     \immediate\write\refmac{\def\noexpand#1{\the\refno}}
   \global\advance\refno by1}
\def\semi{;\hfil\noexpand\break ^^J}
\def\nl{\hfil\noexpand\break ^^J}
\def\refn#1#2{\nref#1{#2}}

\def\up#1{$^{[#1]}$}

\def\jmp#1#2#3{{\it J. Math. Phys.} {\bf {#1}} (19{#2}) #3}

\def\pl#1#2#3{{\it Phys. Lett.} {\bf {#1}B} (19{#2}) #3}
\def\np#1#2#3{{\it Nucl. Phys.} {\bf B{#1}} (19{#2}) #3}
\def\npps#1#2#3{{\it Nucl. Phys. (Proc. Suppl.\/)}
{\bf B{#1}} (19{#2}) #3}

\def\prD#1#2#3{{\it Phys. Rev.} {\bf D{#1}} (19{#2}) #3}

\def\prp#1#2#3{{\it Phys. Rep.} {\bf {#1}C} (19{#2}) #3}

\def\zpC#1#2#3{{\it Z. Phys.} {\bf C{#1}} (19{#2}) #3}


\def\d{\delta}

\def\v#1{v_{_{(#1)}}}

\def\bra#1{\langle#1\vert}
\def\ket#1{\vert#1\rangle}
\def\pa{\partial}

\pageheight{24cm}
\pagewidth{15.5cm}
\magnification \magstep1
\voffset=8truemm
\baselineskip=16pt
\parskip=5pt plus 1pt minus 1pt


\secno=0

{\eightpoint
\refn\PAPER{M. Lavelle and D. McMullan, {\sl Constituent Quarks from QCD},
 Barcelona/Plymouth preprint UAB-FT-369/PLY-MS-95-03}
\refn\JACKIW{{R.\ Jackiw, {\sl Topological  Investigations of
Quantized Gauged Theories,} in {\sl Current Algebra and Anomalies},
ed.'s S.B.\ Treiman et al.
(Princeton University Press, Princeton, 1985)}}
\refn\DAVID{D.\ McMullan, \jmp{28}{87}{428}}
\refn\PANP{M. Lavelle and D. McMullan, \pl{329}{94}{68}}
\refn\DIRAC{P.A.M. Dirac, \lqq Principles of Quantum Mechanics\rqq,
(OUP, Oxford, 1958), page 302}
\refn\GRIBOV{V.N. Gribov, \np{139}{78}{1}}
\refn\FERMI{E. Fermi, {\it Atti della Reale Accademia Nazionale dei
Lincei}, {\bf 12} {(1930)} {431}}
\refn\HAAG{R.\ Haag, {\sl Local Quantum Physics}, (Springer-Verlag,
Berlin, Heidelberg, 1993)}
\refn\BUCH{D.\ Buchholz, \pl{174}{86}{331}}
\refn\PR{P.\ Pascual and E.\ de Rafael, \zpC{12}{82}{12}}
\refn\WILSON{K.G.\ Wilson, \prD{10}{74}{2445}}
\refn\USSSB{M.\ Lavelle and D.\ McMullan, \pl{347}{95}{89}}
\refn\NEUBERT{M.\ Neubert, \prp{245}{94}{259}}
\refn\LATTGRIBOV{P. de Forcrand and J.E. Hetrick, \npps{42}{95}{861}}
}
%
%
\leftline{\bf FINAL REVISED VERSION}
\rightline {UAB-FT-375}
\rightline {PLY-MS-95-06}
\vskip 40pt
\centerline{\bigbold THE COLOUR OF QUARKS}
\vskip 30pt
\centerline{\smc Martin Lavelle{\hbox {$^1$}}
and  David McMullan{\hbox {$^2$}}}
\vskip 15pt
{\baselineskip 12pt \centerline{\null$^1$Grup de F\'\i sica Te\`orica
and IFAE}
\centerline{Edificio Cn}
\centerline{Universitat Aut\'onoma de Barcelona}
\centerline{E-08193 Bellaterra (Barcelona)}
\centerline{Spain}
\centerline{email: lavelle@ifae.es}
\vskip 13pt
\centerline{\null$^{2}$School of Mathematics and Statistics}
\centerline{University of Plymouth}
\centerline{Drake Circus, Plymouth, Devon PL4 8AA}
\centerline{U.K.}
\centerline{email: d.mcmullan@plymouth.ac.uk}}
\vskip 7truemm
\vskip 40pt
{\baselineskip=13pt\parindent=0.58in\narrower\ni{\bf Abstract}\quad
It is shown that colour can only be defined on gauge invariant states.
Since the ability to associate colour with constituent quarks is an
integral part of the constituent quark model, this means that,
if we want to extract constituent quarks from QCD,
we need to dress Lagrangian quarks with gluons so that the result
is gauge invariant.
We further prove that gauge fixings can be used to construct such
dressings.
Gauge invariant dressed quark states are presented and a direct approach
to the interquark potential is discussed. Some further aspects of
dressing quarks are
briefly discussed.
\par}

\vfill\eject
\noindent In the naive quark model colour had to be introduced to
preserve Fermi statistics for the quarks. As such quarks were
identified as spin 1/2 particles with an SU(3) colour index. This
successful, phenomenological approach has now been subsumed into
Quantum Chromodynamics (QCD). Here the Lagrangian fermion fields
also carry an SU(3) index which it is natural to view as colour and
hence the QCD quarks are generally identified with those of the
quark model. Of course it is usually understood that this
identification needs some refinement --- the Lagrangian fermions are
not gauge invariant and the masses of the
(light) Lagrangian quarks are about one hundredth of their
constituent counterparts. However, we will see in this letter
that the Lagrangian
fermions do not even have a well defined colour. Hence the construction
of constituent quarks from QCD\up{\PAPER} must start with a better
understanding of the colour of quarks in QCD.

The structure of this letter is as follows. We first
recall that the QCD
colour charge is not gauge invariant. We then demonstrate the important
property that its action on gauge invariant states {\it is\/} gauge
invariant. It is then shown that gauge
fixings may be used to construct dressings with whose help gauge
invariant dressed quarks may be built up. We will thus show that
dressed quarks with well defined colour quantum numbers can be
constructed in QCD. The rest of the letter is given over to describing
some applications of such dressed fields. In particular we sketch a
new, simple approach to the QCD static interquark potential which
is based upon dressed quarks.
\bigskip
Let us now demonstrate our claim
that the Lagrangian quarks of QCD  have an
ill-defined colour. The colour charge, $Q=Q^aT^a$, corresponds to
the generator of rigid gauge transformations:
$$
Q=\int d^3x\,\(J^0_a(x)-f_{abc}E^b_i(x)A^c_i(x)\)T^a
\,,
\eqn\charge
$$
where $J^0_a=-i\psi^{\dag} T^a\psi$ is the matter charge density and
$E_i^a=-F^a_{0i}$ is the chromoelectric field conjugate to the vector
potential, $A_i^a$
($i=1,2,3$). From this we see that if the structure constants are zero,
as in QED, then the matter charge density is the sole contributor to the
electric charge. However, in a non-abelian theory like QCD the charge
has an additional gluonic component, which reflects the well-known
fact that gluons carry  colour. Unfortunately the colour charge differs
from the electric (abelian) one in that it is not gauge invariant. This
obscures any physical interpretation of the colour of a quark or gluon.

It might now be thought that this problem is of little import since all
we see in detectors are colourless hadrons. However, the
success of the quark model teaches us that hadrons to a good
approximation contain constituent objects which obey Fermi statistics.
Thus, for QCD to describe this physics, there has to be a region
where colour can be well defined and associated with constituent
quarks.

To determine this region we recall that the gauge invariance of QCD
implies that not all fields are physical and that it is an example of a
constrained theory. Now it
is known\note{For further details see Ref.\ \JACKIW; for a
discussion of the equivalence of the BRST formalism and this
characterisation of physical states in terms of
Gauss' law see Ref.\ \DAVID.}\ that
physical states may be
characterized by the condition, $G^a\ket{\rm phys}=0$; where $G^a$,
is just the non-abelian generalisation of Gauss' law
$$
G^a(x)=-\frac1g\(D_iE_i\)^a(x)+J^a_0
\,.
\eqn\gauB
$$
As with its abelian counterpart, (\gauB) generates local gauge
transformations.
On physical states the colour charge may now,
easily with the help of (\gauB), be reduced
to the total divergence
$$
Q\ket{{\rm phys}}=\frac1g\int d^3x\,\pa_iE_i\ket{{\rm phys}}
\,.
\eqn\physcharge
$$
Under a local gauge transformation we have $E_i(x)\to U^{-1}(x) E_i(x)
U(x)$. Restricting the transformation $U$ so that at spatial infinity
it reduces to unity, we see\up{\PAPER} that the charge is now invariant
on such locally gauge invariant states. Let us restate this important
result: states invariant under local gauge transformations
have well-defined colour charges.
Since the Lagrangian fermions of QCD are not gauge invariant, Gauss'
law cannot be applied and so
this argument does not apply to them. We have learnt that any
QCD description of constituent quarks must be so invariant.

The question that we now have to address is how can we construct such a
description? Clearly we have to somehow dress the quarks with glue in
such a way that the result is gauge invariant. We now prove that any
gauge fixing may be used to manufacture a dressing such that gauge
invariant quarks may be constructed. Recall that under a gauge
transformation the vector potential, $A_i$, transforms into
$$
A_i^U=U^{-1}A_iU+\frac1gU^{-1}\pa_i U
\,.
$$
Note that it follows from this that $\(A_i^{U_1}\)^{U_2}
=A_i^{U_1U_2}$. Now given some gauge fixing
condition, $\chi(A)$, there must exist an $A$-dependent
gauge transformation, which we
choose to call $h$, such that for an arbitrary field $A$,
$\chi(A^h)=0$. We now want to see how $h$
transforms under gauge transformations. Consider therefore
the gauge related potential,
$A^U$. The gauge transformed $h$ now satisfies $\chi((A^U)^{h^U})=0$.
If $\chi$ is a good gauge fixing the transformation from any vector
potential to the surface gauge choice must be unique.
This implies that under a gauge transformation we have $Uh^U=h$. Armed
with this we can now demonstrate
that such $h$'s can be used to construct dressings. The combination
$$
\psi_{\rm dressed}(x) = h^{-1}(x)\psi(x)\,,
\eqn\psiH
$$
is gauge invariant since
$$
\eqalign{
\psi_{\rm dressed}\to\psi_{\rm dressed}^U= &
h^{-1}(x)U(x)U^{-1}(x)\psi(x)\cr
=&\psi_{\rm dressed}(x)\,.
}
\eqn\no
$$
This completes the proof
that gauge fixings may be used to build up dressings. We note that
its converse, i.e., that dressings may be used to construct gauge
fixings, was proven in Ref.\ \PANP. The dressings which correspond to
different gauge fixings will, of course, have distinct physical
meanings.
As a simple example of this consider the QED case, where the $h$ which
transforms a field into Coulomb gauge, i.e., $\chi(A)=\pa_iA_i$,
is $h_{\rm c}=
\exp(-ie\partial_i A_i/\nabla^2)$. This gauge fixing may be used to
generate the following gauge invariant description of a dressed
electron
$$
\psi_{\rm c}(x)=\exp\(\frac{ie\pa_iA_i}{\nabla^2}(x)\)\psi(x)
\,.
\eqn\static
$$
This dressed field was argued in Ref.\ \DIRAC\
to correspond to a static charge and its propagator was proven in
Ref.\ \PAPER\ to be infra-red finite if the static mass shell
renormalisation point, $p=(m,\vec0)$ is used. Other dressings will
correspond to different physical situations.

In Ref.\ \PANP\ we presented the extension of (\static) to perturbative
QCD up to order $g^2$. In general we do not know the transformation
into Coulomb gauge in non-abelian theories and indeed
non-perturbatively the Gribov ambiguity\up{\GRIBOV} tells us that there
is no such unique transformation. Perturbatively, however, we can
rotate into Coulomb gauge and one finds\up{\PAPER} to order $g^3$
$$
\psi_{\rm c}(x)= \exp\left( g\v1+g^2\v2+g^3\v3\right)\psi(x) +O(g^4)
\,,
\eqn\threedress
$$
where
$$
\v1=\frac{\pa_jA_j}{\nabla^2}\,,\qquad
\v2=\frac{\pa_j}{\nabla^2}\bigl([\v1,A_j]+\tfrac12
[\pa_j\v1,\v1]\bigr)
\,,
\eqn\dressone
$$
and
$$
\eqalign{
\v3=\frac{\pa_j}{\nabla^2}\Bigl([\v2,A_j]+\tfrac12[\v1,[\v1,A_j]]+
&\tfrac12[\pa_j\v1,\v2]+
\tfrac12[\pa_j\v2,\v1]\cr
&-\tfrac16[\v1,[\v1,\pa_j\v1]]\Bigr)\,,
}
\eqn\dressthree
$$
and the commutators are the SU(3) ones. This may be systematically
extended to higher orders\up{\PANP, \PAPER}. We recall from the above
that such dressed quarks have well defined colour
charges and that they may be combined at separate points, just as in
the quark model, to form colourless hadrons.

There is a great deal of literature on charged states in gauge theories
(see, e.g., Ref.'s \FERMI--\BUCH).
It has been shown that they are non-local and non-covariant and the
above dressing indeed displays the expected non-locality
and non-covariance. This does {\it not\/} mean that standard
field theoretic techniques cannot be used. In particular the one loop,
perturbatively dressed fermion propagator can, we stress again,
be renormalised on-shell
resulting in the conclusion that this dressing is that of a static
colour charge\up{\PAPER, \DIRAC}. Furthermore QCD vacuum effects, in
the shape of the quark condensate, can be incorporated into the
propagator. This yields, in the deep Euclidean region,
a gauge invariant, running mass term in the
dressed propagator\up{\PR,\PAPER},
which indicates how one might obtain a constituent quark mass for
dressed charges.

Having now proven that only gauge invariant dressed quarks have well
defined colour charges and having shown that gauge fixings may be used
to construct dressings in a rather direct way, we wish to conclude this
letter by describing some of the more obvious applications of the
dressed quarks. More detailed
discussions of these points may be found in Ref.\ \PAPER.

The interaction between two static colour charges is generally
investigated in terms of Wilson loops\up{\WILSON}. We can, however,
more directly recover the interquark potential by calculating the
energy for a system of two dressed, static quarks as a function of
their separation. Writing the dressed quark as
$\psi_{\rm c}(y)= e^{v_{\rm c}(y)}\psi(y)$, and taking for static
fields the Hamiltonian, $H=\frac12{\rm tr}\int d^3z E^2_i$,
one finds that the
interquark potential, $V(y-y')$, corresponding to a quark and
anti-quark at positions $y$ and $y'$ is
$$
V(y-y')=-\int d^3x\, {\rm tr}\,\bra0[E^a_i(x),v_{\rm c}(y)]
[E^a_i(x),v_{\rm c}(y')]\ket0\,,\eqn\potential
$$
where the trace is over the (anti-Hermitian) colour matrices and
the commutators are  the canonical equal time ones
between the chromoelectric
field and the potential:
$$
[E^a_i(x),A_j^b(y)]=iZ_3^{-1}\d_{ij}\d_{ab}\d(x-y)\,.\eqn\ccr
$$
with $Z_3$ the gluon wave function renormalization constant.
A perturbative expansion for the potential now follows directly from
the perturbative description of the static dressing (\threedress),
and to order $g^2$ we immediately recover the usual Coulomb interaction.
Note that knowing the dressing to order $g^3$ is sufficient to
determine the potential to order $g^4$.
The form of the explicit expressions (\dressone)
and (\dressthree) for the dressing to this order shows that the $g^4$
contribution to the potential only involves  projections of the
free gluon propagator along with the known results for $Z_3$ to
this order.
This should be contrasted with the Wilson
loop approach where a one loop perturbative calculation must be
inserted into the contour integral around the static loop. In general
it may be seen\up{\PAPER} that this more direct calculation
involves additional calculations
to one loop less than the standard method and, of course,
the contour integral
and associated large T (length of the Wilson
loop in the time direction) limit
are not present at all. This is because of our
knowledge of the dressing and it makes higher order calculations of the
potential significantly easier.

The dressings we have explicitly considered above have corresponded to
static charges and been based upon transforming into Coulomb gauge.
The use of a different transformation permits us
to build up dressings for
non-static charges. To lowest order in $g$ we find for small velocity,
${\bold v}$,  that\up{\PAPER}
$$
\psi_{\bold v}(x)=
\exp\left(
g\frac{\pa_jA_j+ v_i E_i}{\nabla^2}
\right)\psi(x)
\,,
\eqn\mover
$$
where ${\bold v}$ is the velocity of the charge. Following Ref.\
\DIRAC\ it may be seen that the electric and magnetic fields
associated with (\mover) are indeed those of a moving charge up to
order ${\bold v}$.
Dressings appropriate to relativistic quarks may be found
elsewhere\up{\PAPER}. It would be interesting to study the implications
of such
dressings for moving fields for both jet physics and the heavy quark
effective theory\up{\NEUBERT}.

The connection between gauge fixings and dressings has a further
important consequence in QCD. Although, as we have seen, perturbative
dressings can be constructed corresponding to a wide class of quark
states, a non-perturbative dressing would require a non-perturbative
gauge fixing. In QED and in spontaneously broken non-abelian gauge
theories such gauge fixings may be built up\up{\USSSB}.  However, in
QCD the Gribov ambiguity\up{\GRIBOV} precludes the existence of a
non-perturbative dressing. This means that no asymptotic quark state
may be constructed, which is consistent with experimental results.

This approach opens\up{\PAPER} various routes to the determination of
hadronic scales from QCD. One can, for example, construct a Wilson loop
type operator using the dressed quark fields and see where the Gribov
ambiguity first appears. It should be noted here that the Gribov
ambiguity on the lattice seems to be  additionally complicated by
lattice artifacts\up{\LATTGRIBOV}.
Alternatively one can directly calculate the
interquark potential, as discussed above, and see where  it first
becomes energetically favourable for the dressing of a mesonic state
to no longer factorize into a product of constituent dressings.
Although these ideas will entail non-perturbative calculations, we feel
that new approaches in a field as thorny as confinement are to be
welcomed and that these suggestions may be eventually fruitful.

In conclusion we have seen that the colour charge is only defined on
gauge invariant states. Such quark states have been constructed
perturbatively and we have seen that there is a non-perturbative
obstruction to their formation.

\bigskip
\ni {\bf Acknowledgements:} MJL thanks project CICYT-AEN95-0815 for
support and the University of Plymouth for their hospitality.
\bigskip
\bigskip
\bigskip

  \immediate\closeout\reffile
  \centerline{{\bf References}}\bigskip\eightpoint\frenchspacing%
  \input refs.tmp\vfill\eject\nonfrenchspacing

\bye